\documentclass[11pt, twocolumn]{article} 

\usepackage{amsmath}
\usepackage{amssymb}
\usepackage{lipsum}
\usepackage{mathrsfs}

\usepackage{graphicx}
\usepackage{bm}
\usepackage[T1]{fontenc}
\usepackage{url}
\usepackage{subcaption}
\usepackage{rotating}
\usepackage{soul}
\usepackage{scalefnt}
\usepackage[hidelinks]{hyperref}
\usepackage[utf8]{inputenc}
\usepackage{enumerate}
\usepackage{float}
\usepackage{authblk}
\usepackage{url}
\usepackage{amsthm}

\usepackage{color}
\definecolor{redcolor}{rgb}{1.0,0.,0.}

\title{Modeling the Perspectives for Scientific Advancement}

\author[1]{Eric K. Tokuda}
\author[2]{Cesar H. Comin}
\author[1]{Luciano da F. Costa}

\affil[1]{S\~ao Carlos Institute of Physics, University of S\~ao Paulo, S\~ao Carlos, SP, Brazil}
\affil[2]{Department of Computer Science, Federal University of S\~ao Carlos, S\~ao Carlos, SP, Brazil}
\date{}

\begin{document}

\twocolumn[
  \begin{@twocolumnfalse}
    \maketitle
    \begin{abstract}

The development of science constitutes itself an important subject of scientific investigation.Indeed, better knowledge about this intricate dynamical system can provide subsidies for enhancing the manners in which science progresses.  Recently, a network science-based approach was reported aimed at characterizing and studying the prospects for scientific advancement assuming that new pieces of knowledge are incorporated in a uniformly random manner. A surprising result was reported in the sense that quite similar advancements were observe for both Erd\H{o}s–R\'enyi (ER) and Barab\'asi–Albert (BA) knowledge networks. In the present work, we develop a systematic complementation of that preliminary investigation, considering an additional network model, the random geometric graph (GR) as well as several other manners to incorporating knowledge, namely preferential to node degree, preferential to closer or adjacent nodes, as well as taking into account the betweenness centrality of the unknown nodes. Several interesting results were obtained and discussed, including: 
the uniform strategy led to the best expansion in the GR model, 
the results of the degree and betweenness expansion for the ER and GR models were similar to those obtained using uniformity method in those same models.  Surprisingly, the BA model led to 
just a slightly faster expansion than in the uniform case.  Though qualitatively similar, 
strategies based on the degree and betweenness yielded distinct results as a consequence of
them not being linearly related.
    \end{abstract}
\vspace{2em}
  \end{@twocolumnfalse}
]

\section{Introduction}  \label{s:introduction}

Science can be approached and studied as a complex system
whose dynamics take place over space and time.  Along centuries, or even
millenia, science has progressed from relatively simple concepts --
typically covered as a single discipline -- to a universe of 
specialized areas involving highly sophisticated models and
theories.

At any given time, great attention is often given to identifying 
the \emph{prospects for scientific advancement}, which are
understood in a somewhat subjective manner as corresponding to the
number and importance of the potential developments that can be
achieved in the light of the existing knowledge.  The accurate identification
of these prospects is of particular interest because they it help
focusing and prioritizing research and funding resources along specific 
venues that are more related or of particular relevance in a
given time and location.  

The impressive development of the area known as network science (e.g.~\cite{newman2003structure})
(roughly speaking, the study of complex networks) was to a great 
extent a consequence of the generality of graphs (also known networks) 
for representing virtually every discrete system (e.g.~\cite{costa2011analyzing}), ranging from the Internet to molecules interaction.  Indeed, it is
even possible to represent the scientific body of knowledge as a 
complex network, e.g.~by representing portions of knowledge as nodes
while expressing the respective relationships in terms of edges
(e.g.~\cite{da2006learning, shi2015weaving}).

Such representations of scientific knowledge in terms of networks
have paved the way to a number of interesting possibilities in 
scientometrics, the science that studies science (e.g.~\cite{foster2015tradition, rzhetsky2015choosing, silva2010identifying, amancio2012using}
For instance, it becomes possible to model scientific advancement
in terms of different types of random walks along these representations 
(e.g.~\cite{lima2018dynamics,batista2010knowledge}).  It also becomes possible to model and
simulate the transmission of information and knowledge (e.g.~\cite{de2017knowledge, de2019connecting}).

Complex network representations of scientific knowledge also provide
subsidies for the main motivation of the present work,
namely the definition and quantification of the prospects for scientific
advancement.  In a preliminary work~\cite{dacosta2020oncomplexity}, having represented the overall
scientific knowledge as a complex network, a subset of these nodes
was understood as the \emph{nucleus}, representing the currently
known portion of the knowledge.  Then, the prospects, or potential for knowledge advancement  at that particular stage could be objectively 
defined and quantified as deriving from the number of yet unknown nodes that can be accessed from the nucleus. More specifically, the indices $r$ and $s$ were defined. Index $r$ indicates the prospect of knowledge expansion around the nucleus, while index $s$ is associated with a redundancy of the knowledge advancement. 

As described in~\cite{dacosta2020oncomplexity}, it is not only possible to quantify 
the potential for scientific advancement with respect to the current
stage, but also to study how this potential unfolds as the nucleus
is progressively expanded, e.g.~by selecting new node with
uniform random probability.  In those circumstances, a remarkable
result was observed, namely that the prospects do not depend
on the topology of the network (two models with quite different
topological properties, namely  Erdős–Rényi (ER) and Barabási–Albert (BA), were considered in ~\cite{dacosta2020oncomplexity}.

One of the objectives in~\cite{dacosta2020oncomplexity} was to study the change of $r$ with the size $c$ of the nucleus. It was shown that a typical curve has two parts: it increases with the increase of $c$, achieves a maximum $r_{max}$ at $c_{rmax}$ and then steadily decreases with $c$ until $r=0$. When a curve has a lower $c_{rmax}$ compared to another, it means that it achieves the maximum value of $r$ for a smaller value of $c$. That is to say that it achieves the optimal value of the potential knowledge expansion with a smaller nucleus size.

The surprising result that quite similar prospects were observed
for varying topologies was understood to be a consequence of the
uniformly random choice of nodes to be incorporated into the nucleus,
and it was foreseen that other strategies of nucleus expansion could
lead to distinct results.

The present work resumes and expands those investigations.
Not only we consider additional complex networks models,
namely the random geometric graph (GR) but more importantly we study other manners of expanding the nucleus preferentially in terms of degree, 
distance, adjacency and betweenness centrality. Several interesting results are obtained, including the surprising effectiveness of the uniform expansion strategy on the GR model, the fact that the degree strategy was just slightly more efficient in the BA model than the uniform strategy, as well as the large variation in the results observed on the GR model depending on the chosen strategy.

This work starts by presenting the adopted data and basic methods,
including models and properties of complex networks.  Then, the main
framework employed for modeling the unfolding of the prospects of 
scientific advancement is described, including the measurements that
are used for respective quantification.  The main obtained results are then
respectively presented and discussed.

\section{Data and Methods}
\label{sec:method}

The ER model is among the most frequently studied network type. One of the proposed procedures to obtain these networks consists in starting with an isolated set of vertices and then gradually adding edges with uniform probability among all the possible edges. In this construction scheme, given a fixed number of vertices, and a desired number of edges, all graphs are equally probable. 

The GR model is another category of random graphs in which the specification of the position of the vertices in a given metric space is performed in random manner. The edges are, then, deterministically created by linking vertices at most a given distance apart. This procedure leads to the generation of clusters of vertices with high modularity, which do not appear in the ER model.

Despite the interest motivated by the simplicity of these networks, there are relatively few real-world
structures that can be respectively modeled. In particular, it has been noted that real networks have uneven degree distribution. In this context,  the BA model was proposed as being capable of generating power law degree distributions. Such networks are also called scale-free (or scale invariant) networks. The construction of a BA model consists in a growth model following a preferential attachment rule, which is linear to the degree of the candidate vertices. This \emph{rich gets richer} phenomenon principle leads to the generation of a few highly connected nodes - the \emph{hubs}.

The complexity characterizing many network models motivated the development of a myriad of measurements. One of most basic measurements is the vertex degree, which in an undirected network represents the number of connections of the vertex. Network topologies can be initially compared based on their degree probability distribution over the network. In different applications, such as epidemics and urbanism, it is often useful to identify the \emph{most important} vertices . The concept of \emph{importance} is application-dependent, giving rise to multiple approaches to centrality characterization. The betweenness centrality is one of these measurements, based on the shortest path measurement.  More specifically, it counts the the number of shortest paths that pass through each vertex.

\section{The Proposed Framework}

Being composed of a set of concepts and interrelationships, knowledge
can be effectively represented in terms of a network or a graph.  
Having decided how the portions of knowledge are to be assigned to
nodes, which involves choosing a respective level of detail, one
maps each of these portions as a node, while interrelationships are
represented in terms of respective edges.  There are several types of
interrelationships that can be considered, including pre-requisites
(e.g.~one need first to learn topic A before proceeding to topic B),
application of theorems and results, or simply a citation of words
associated to the chosen topics.  

Once knowledge is organized in this manner, we can represent its
portion that is already known in terms of the respective nodes, which
gives rise to the concept of \emph{nucleus}.  Though other approaches can be
adopted, in this work we consider all the interconnections between
the known nodes.  Therefore, all nodes other than
those belonging to the nucleus correspond to pieces of the knowledge
that are not yet known, being associated to the perspectives for 
scientific advancement.  The progress of our understanding of the
knowledge represented in the overall network can therefore be
understood as the progressive incorporation of previously unknown
nodes into the nucleus, which is consequently expanded until all
nodes are encompassed.

We propose to study the dynamics of the prospect for scientific
advancement as we increase the nucleus by using two different measurements, \emph{r} and \emph{s}, as previously proposed in~\cite{dacosta2020oncomplexity}. The measurement \emph{r}, defined as the ratio between the number of nodes adjacent to the nucleus not yet visited (\emph{n}) and the total number of nodes (N),
i.e.:
\begin{equation}
  r=\frac{n}{N},
\end{equation}

This measurement can be seen as a relative indication of the prospect for scientific advancement. 
The other proposed measurement, \emph{s}, defined as the ratio between the number of n and the number of \emph{edges} that connect the nucleus to these adjacent nodes not yet visited, i.e.:
\begin{equation}
    s=\frac{n}{e}
\end{equation} 

\begin{figure}[ht]
  \centering
  \includegraphics[width=.42\textwidth]{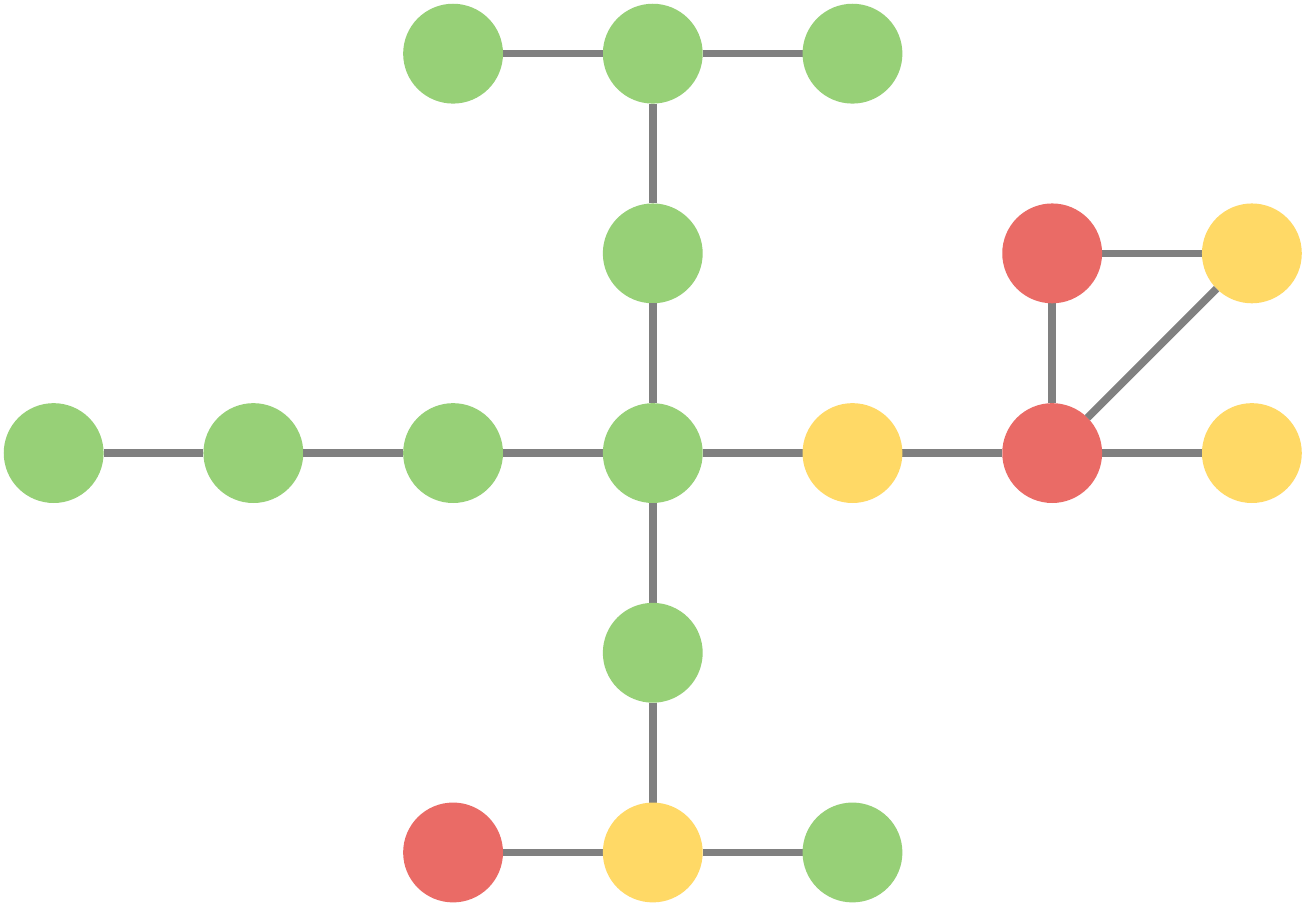}
  \caption{Calculation of the proposed measurements in an example network. The green nodes represent knowledge not yet discovered, the red ones represent the nucleus, and the yellow nodes represent the knowledge adjacent to the nucleus. In this example, N=17, n=4 and e=6 and hence, $r=4/17$ and $s=4/5$.}
  \label{fig:diagramrs1}
\end{figure}

In Figure~\ref{fig:diagramrs1}, the nucleus is represented by the set of the $n$ red nodes. The 
yellow nodes represent the adjacent nodes and the green ones represent the non-discovered nodes. 
The edges between the nucleus and the adjacent nodes is denoted by $e$. In the example shown, the nucleus is composed of three vertices and
$r=n/N=4/17$ and $s=n/e=4/5$.

\begin{figure}[ht]
  \centering
            \centering
    \begin{subfigure}[b]{.21\textwidth}
        \includegraphics[width=\textwidth]{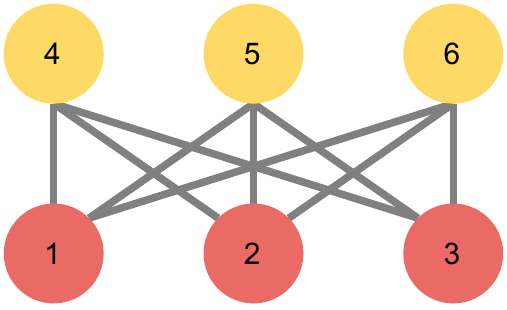}
        \caption{}
    \end{subfigure} \qquad
    \begin{subfigure}[b]{.21\textwidth}
        \includegraphics[width=\textwidth]{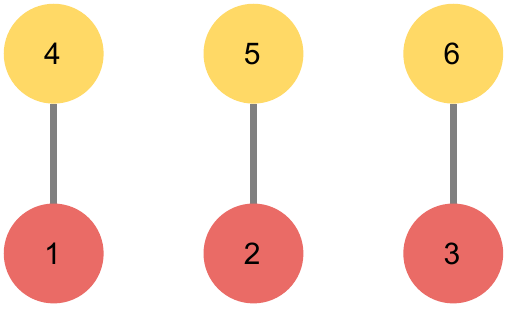}
        \caption{}
    \end{subfigure}
  \caption{An illustration of how nodes adjacent to nucleus can be reached
  in different manners, expressed in terms of measurement $s$. 
  In (a), nodes 4, 5 and 6 can be found from nodes 1, 2, and 3, therefore
  they have $s = 3/9 = 1/3$.  Contrariwise, in (b) we have $s = 3/3 = 1$.}
  \label{fig:diagramrs2}
\end{figure}

However, not all these unknown nodes are likely to be equally 
accessible from nodes in the nucleus, which gives rise to several 
distinct possibilities of expanding the nucleus.  Figure~\ref{fig:diagramrs2}
illustrates this effect in terms of the measurement $s$.

The most elementary possibility, already considered in~\cite{dacosta2020oncomplexity} 
consists of choosing unknown nodes in uniformly random manner for respective
incorporation into the nucleus.  Another interesting possibility
is to choose among the unknown nodes with probability proportional
to their respective degree, reflecting a tendency that pieces of
knowledge with more interconnections are more likely to be discovered.

It is also interesting to consider expansion strategies relying on
the distance or adjacency between nodes.  For instance, nodes
that are closer to those belonging to the nucleus can be assumed to
be more likely to be learned, according to a respective probabilistic
model.  Alternatively, unknown nodes that are adjacent to the nucleus
(i.e.~directly connected to at least one of the nucleus nodes) can 
be chosen as the next pieces of knowledge to be discovered.

It is also interesting to consider the \emph{betweenness centrality}
of the network vertices.  In this manner, it is possible to
derive a nucleus expansion strategy that takes into account the
number of shortest paths going through each node, taking into 
account the possible interrelationships between the pieces of
knowledge in the overall network.

In this work, we adopted all the above motivated methodologies for 
progressively extending the nucleus.  More specifically, we have: 
(i) unknown nodes are chosen with uniform probability; (ii) preferential to 
the vertex degree; (iii) with probability $p(v) = \alpha \cdot exp(-\beta d)$, 
where $d$ is the distance; (iv) the unknown nodes connected (adjacent) to 
the nucleus are chosen with uniform random probability; and (v) 
preferential to the betweenness centrality.

\begin{figure}[ht]
        \centering
\includegraphics[width=.49\textwidth]{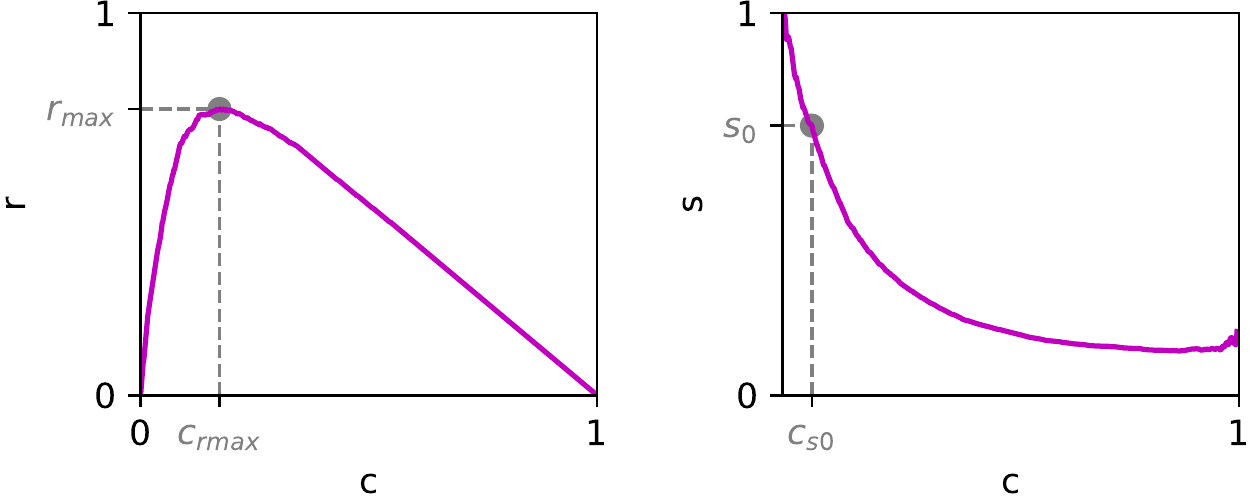}
        \caption{Measurement of \emph{r} and \emph{s} from an actual experiment. 
        On the left and on the right the variation of \emph{r} and \emph{s}, 
        respectively, as we expand the relative size of the nucleus (\emph{c}). 
        The index \emph{r} achieves a maximum value at the point
        ($c_{rmax}$, $r_{max}$)  and \emph{s} decreases to $1/\sqrt{2}$  
        at ($c_{s0}, s_0$).}
    \label{fig:rmaxcs0}
\end{figure}

Figure~\ref{fig:rmaxcs0} illustrates the signatures of the measurements \emph{r}
and \emph{s} typically obtained in our simulations.  As every signature
of \emph{r} is qualitatively similar, it is enough to characterize these
curves in terms of the two measurements $c_{rmax}$ and $r_{max}$ indicated in Figure~\ref{fig:rmaxcs0}
corresponding to the position where the peak of $r$ is observed and the
value of this peak.

In the case of the signatures obtained for \emph{s}, which are also all
qualitatively similar, we express the intensity of the decrease of 
\emph{s} in terms of the value of \emph{c} where the value of 
\emph{s} falls from 1 to $\sqrt{2}/w \approx 0.707$, as frequently adopted
in physics and engineering.

\section{Results and Discussion}

We performed three main experiments, in which we varied the size and 
average degree of the network according to the following configurations: 
(1) $N= 300$ and  $\left< k \right> = 12$; (2) $N= 300$ and 
$\left< k \right> = 18$; and (3) $N= 700$ and $\left< k \right> = 12$.   

\begin{figure}[ht]
        \centering
        \includegraphics[width=.48\textwidth]{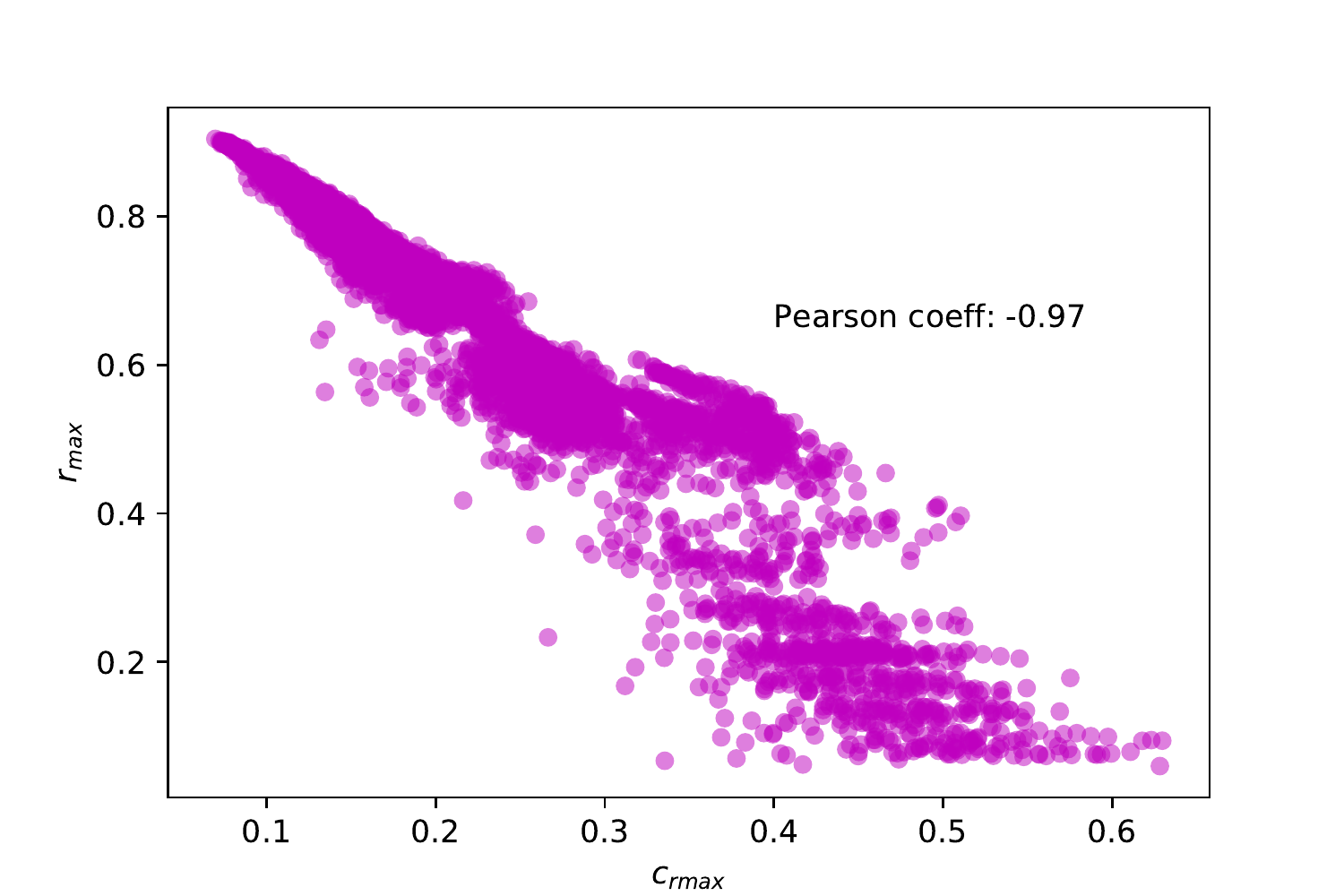}
        \caption{Correlation between the maximum \emph{r} achieved ($r_{max}$) and the
        corresponding \emph{c} value ($c_{rmax}$).  Though a particularly high
        Pearson correlation coefficient is observed, the relationship
        between these two measurements is far from being perfectly linear.}
    \label{fig:corrrmaxcrmax}
\end{figure}

Figure~\ref{fig:corrrmaxcrmax} depicts the scatterplot of $r_{max}$ versus $c_{rmax}$. 
This results indicates that these two measurements are strongly interrelated,
so that our analysis will concentrate on the former measurement (i.e.~$r_{max}$).
The dispersion in Figure~\ref{fig:corrrmaxcrmax} can be observed to be stronger when 
$r_{max}$ is small and $c_{rmax}$ is large.  In other words, the position where 
smaller peaks are observed is more difficult to be predicted from $c_{rmax}$.

Therefore, the following discussion focuses on the measurements $r_{max}$ and $c_{s0}$.

Each of these experiments are respective to the ER, BA, and GR theoretical
models of complex networks, and five strategies for controlling the
expansion of the kernel, namely uniformly random, dilation, preferential to the degree, to the betweenness centrality, and to the distance.

\begin{figure*}[ht]
            \centering
    \begin{subfigure}[b]{.32\textwidth}
        \includegraphics[width=\textwidth]{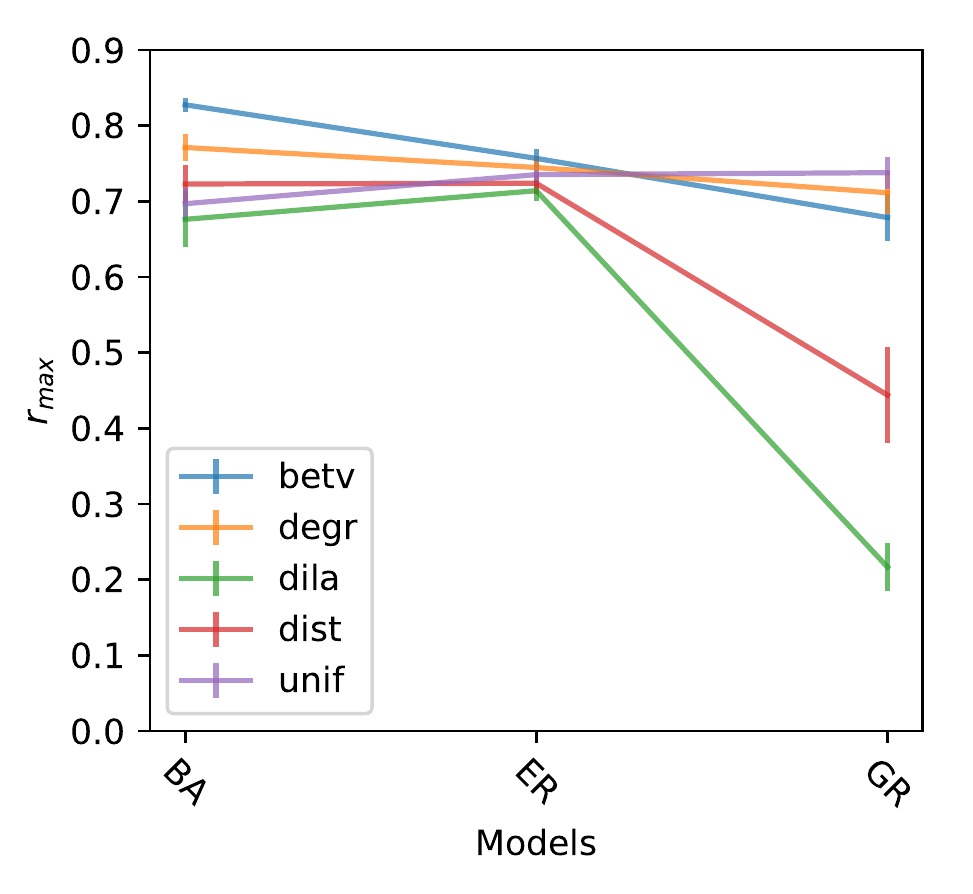}
        \caption{}
    \end{subfigure}
    \begin{subfigure}[b]{.32\textwidth}
        \includegraphics[width=\textwidth]{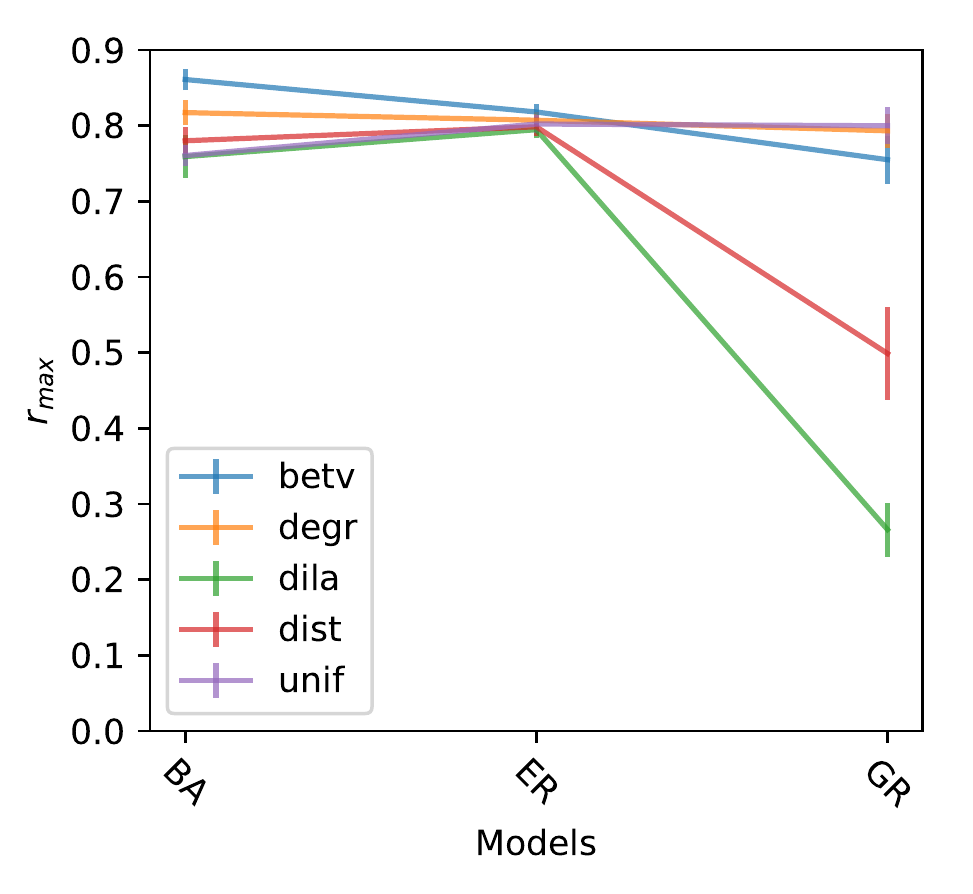}
        \caption{}
    \end{subfigure}
    \begin{subfigure}[b]{.32\textwidth}
        \includegraphics[width=\textwidth]{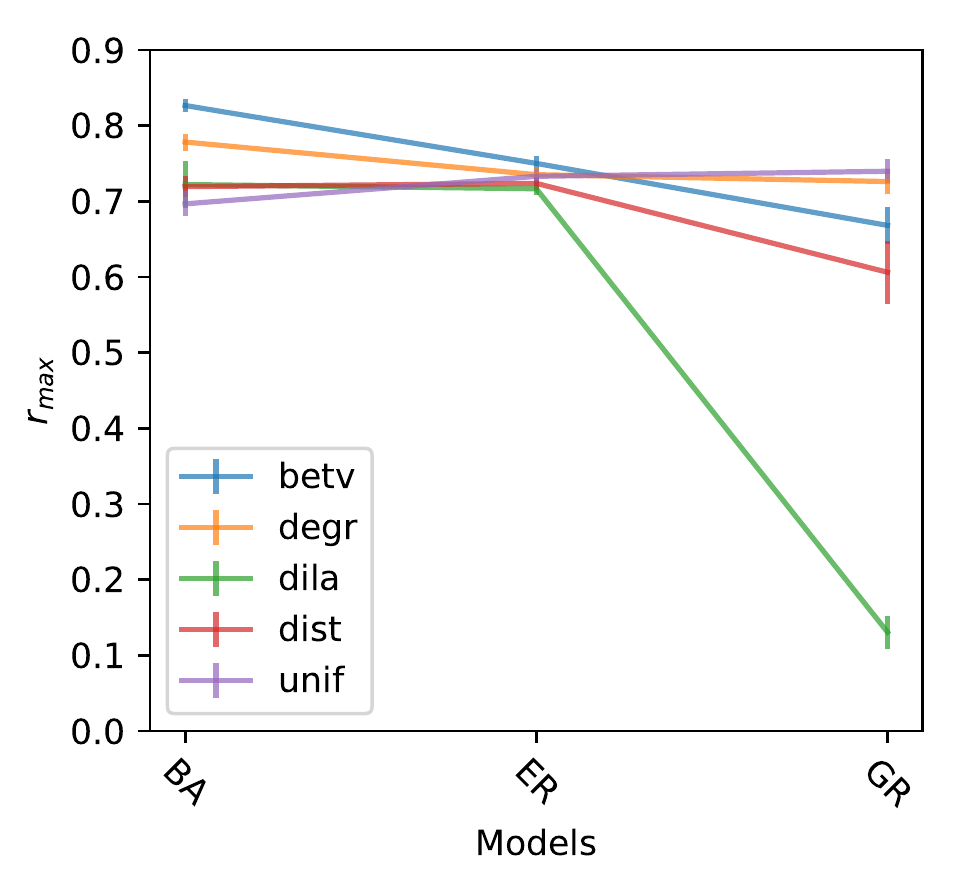}
        \caption{}
    \end{subfigure} \\
    \begin{subfigure}[b]{.32\textwidth}
        \includegraphics[width=\textwidth]{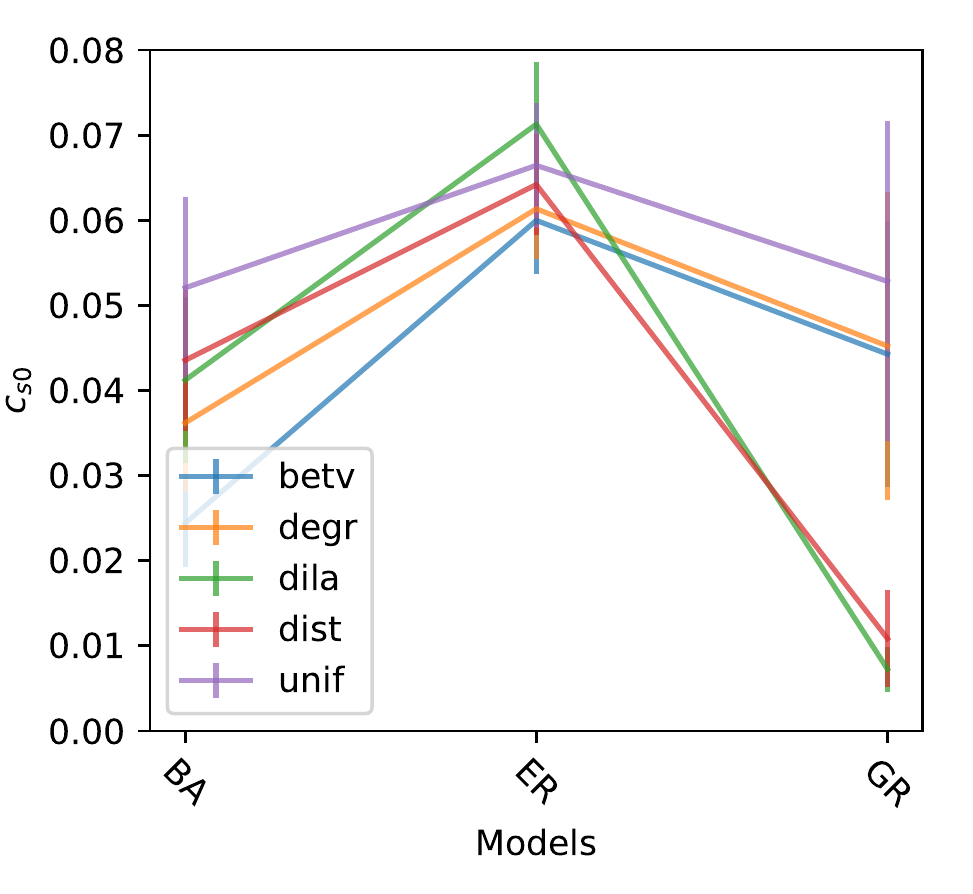}
        \caption{}
    \end{subfigure}
    \begin{subfigure}[b]{.32\textwidth}
        \includegraphics[width=\textwidth]{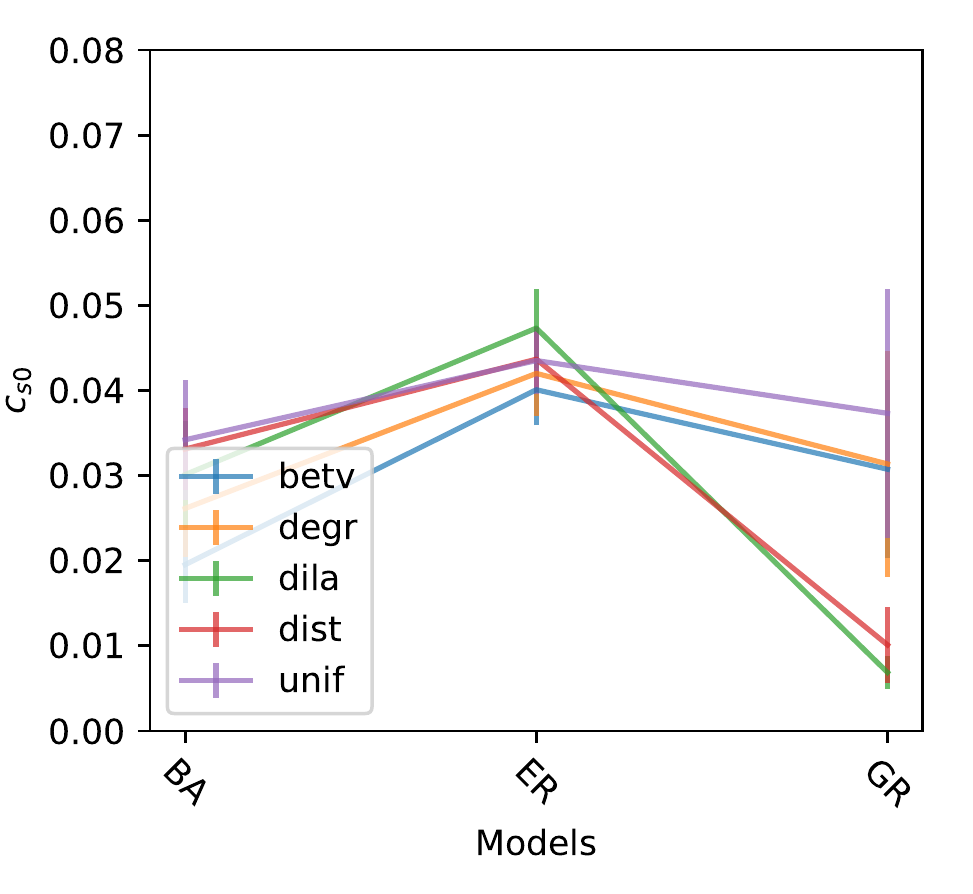}
        \caption{}
    \end{subfigure}
    \begin{subfigure}[b]{.32\textwidth}
        \includegraphics[width=\textwidth]{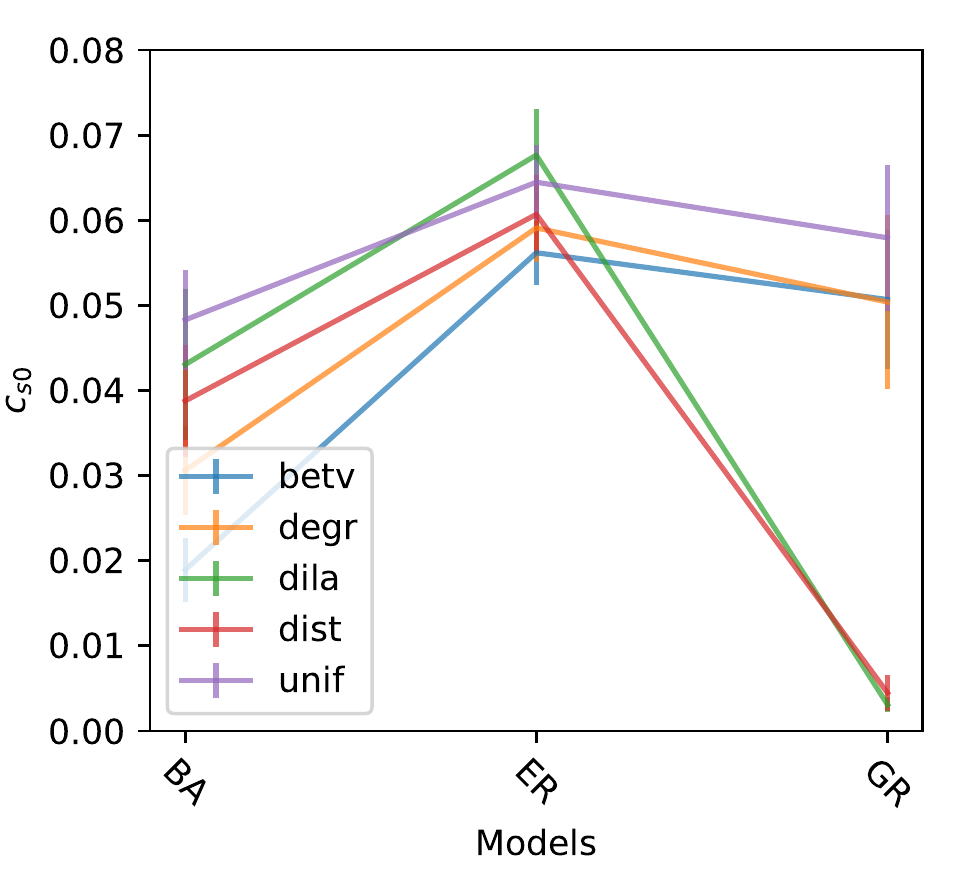}
        \caption{}
    \end{subfigure} 
        \caption{Comparison among the the topologies and the different strategies of increase of the nucleus based on the the two proposed measurements. In the first row and in the second rows consider the measurement r and s, respectively. The columns represent different graph parameters: (300, 12), (300, 18), (700, 12), for the first, second and third columns respectively, with $(n, k)$ representing graphs with $n$ vertices and average degree $k$.}
    \label{fig:resultsall}
\end{figure*}

The results obtained are depicted in Figure~\ref{fig:resultsall}.

The results were found to be similar for different network sizes and average degrees, with the exception of the expansion strategy based on the distance, in which case the $r_{max}$ increases
with the network size for the GR model.  All other results are qualitatively similar
with the network size, and are therefore discussed together in the following

As could be expected, the results for the uniform expansion strategy were similar to those obtained in~\cite{dacosta2020oncomplexity}. It is interesting to note that the uniform method led to the best expansion in the GR model. This model had not been considered in~\cite{dacosta2020oncomplexity}. Regarding the degree expansion strategy, the results for the ER and GR models were similar to those obtained in the uniform case. Surprisingly, the BA model resulted in just a slightly faster expansion than in the uniform case. The results for the distance-based expansion strategy were similar to those obtained for the BA and ER models using the uniform strategy. Contrariwise, the expansion was much slower for the GR model when the distance approach was employed. This contrast between the models is likely due to the difference in diameter between the networks generated by the models. In the case of the ER and BA networks, almost all nodes can be reached from the nuclei after two steps for the considered network sizes. In the GR model a distinct behavior is observed, the knowledge becomes localized around the nuclei, and expands by an amount that is associated with the size of the nuclei periphery.

\begin{figure}[ht]
        \includegraphics[width=.48\textwidth]{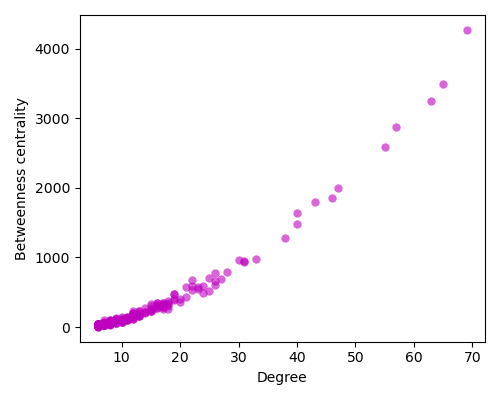}
        \caption{Correlation between the vertex degree and the betweenness centrality for all the experiments with topology BA.}
    \label{fig:corr}
\end{figure}

Regarding the dilation strategy, the results were similar to those obtained by the distance-based expansion, but were more pronounced in the sense that it represented the worst expansion strategy among all models. The results for the betweenness strategy were similar to those obtained by the uniform strategy in the ER and GR models. Interestingly, the betweenness led to the best expansion strategy in the BA model, being even better than the degree approach. This happens because the betweenness grows very fast as the degree of the nodes increase. Actually, it has as superlinear relationship with the degree, as can be seen in Figure~\ref{fig:corr}. This means that hubs will have a disproportionately large preference during the knowledge expansion, leading to a better coverage of the network.

It is important to note that all expansion strategies led to similar results in the ER model. This is because the vast majority of the nodes have similar properties for this model. On the other hand, the results for the GR model were highly dependent on the strategy used. Even though the BA model is known for being heterogeneous~\cite{albert2002statistical}, the GR model had a higher impact on the considered strategies depending on the node property used. This is likely due to the large diameter of the networks generated by the model, which allows highly distinct expansion dynamics according to the knowledge acquisition strategy. The BA model, having small-world properties, does not yield the same variability.

Concerning the measurement $c_{s0}$ shown in Figures~\ref{fig:resultsall}(d)-(f), though the results vary with the network degree, they have similar relative variations according to the network model. 

As far as the absolute variations are concerned, the values in (e) are smaller than in (d) 
because of the increase of average degree, which implies more links to converge on the
nodes adjacent to the nucleus.  No variation is however observed with the increase of
network size from (d) to (f).

Concerning the relative variations, the strategies were found to have similar effects in the
case of the ER models, while larger differences were observed for BA, and even larger changes
for GR.  These effects are analogous to those observed and discussed for the $r_{max}$
measurement. The values for $c_{s0}$ are positively correlated with $r_{max}$ for the BA model, while a negative correlation is observed for the GR model. The values obtained for the ER model are larger than those obtained for the BA and GR. Since the GR model has a large clustering coefficient~\cite{dall2002random}, it is expected that the connections will tend to be more convergent. In the case of the BA model, many nodes in the nucleus will tend to be connected to a hub, lowering the value of $S$. 

The effect of the strategies is analogous to that observed and discussed in the case of
$r_{max}$, with the difference that the strategies based on dilation and distance were
now found to be very similar.

\section{Concluding Remarks}

From its beginnings, science has relied on the identification of prospects 
for subsequent advancements, which can not only contribute to better planning
but also for making the scientific activity more efficient.  Yet, despite such
central importance, the issue of defining and identifying the prospects for
scientific advancement have been approaches in a predominantly subjective manner.

Representing the continuation of a recently reported preliminary approach to this
problem~\cite{dacosta2020oncomplexity}, the current work described a substantially extended investigation on how
the prospects for scientific advancement can be quantified in a more objective
manner and how it can be predicted given network-based representations of bodies
of knowledge.

More specifically, the knowledge is represented as a network by assigning its
components to nodes, while the relationships between these components are 
represented by respective links.  Three distinct models have been considered for
this finality, namely ER, BA, and GR, each being characterized by specific
topological properties.  The current knowledge is understood to correspond to
a subgraph of the overall network, which is called nuclei.  The prospect for
advancement of knowledge can then be defined as corresponding to the number
of nodes that are adjacent to the nuclei.

By adopting several different strategies for expanding the nuclei along time,
it became possible to study how the prospects respectively changed. Several interesting results were obtained and discussed. A surprising result was that the uniform strategy led to a relatively good expansion of the nucleus for all models, when compared to the other strategies. Surprisingly, this strategy was the most effective for the GR model. Another interesting result was that despite the heterogeneity of the BA model, similar results were obtained for all expansion strategies. On the other hand, the results for the GR model varied broadly according to the strategy used. 

The results presented in this work are a first step towards modeling knowledge expansion using complex networks. Furthermore, the proposed dynamics is not limited for modeling knowledge expansion and can be applied to other areas. For instance, an interesting prospect is to quantify the accessibility of cities using the proposed $r$ and $s$ indices on road networks.

\section*{Acknowledgments}

Eric K. Tokuda thanks FAPESP (grant no. 19/01077-3) for financial support. Cesar H. Comin thanks FAPESP (grant no. 18/09125-4) for financial support. Luciano da F. Costa thanks CNPq (grant no. 307085/2018-0) for sponsorship. This work has also been supported by the FAPESP grant 15/22308-2.

\bibliography{main}

\begin{thebibliography}{10}

\bibitem{albert2002statistical}
R{\'e}ka Albert and Albert-L{\'a}szl{\'o} Barab{\'a}si.
\newblock Statistical mechanics of complex networks.
\newblock {\em Reviews of modern physics}, 74(1):47, 2002.

\bibitem{amancio2012using}
Diego~R Amancio, Maria das Gra{\c{c}}as~Volpe Nunes, Osvaldo~N Oliveira~Jr, and
  L~da~F.~Costa.
\newblock Using complex networks concepts to assess approaches for citations in
  scientific papers.
\newblock {\em Scientometrics}, 91(3):827--842, 2012.

\bibitem{batista2010knowledge}
JB~Batista and L~da~F Costa.
\newblock Knowledge acquisition by networks of interacting agents in the
  presence of observation errors.
\newblock {\em Physical Review E}, 82(1):016103, 2010.

\bibitem{costa2011analyzing}
Luciano da~Fontoura Costa, Osvaldo~N Oliveira~Jr, Gonzalo Travieso,
  Francisco~Aparecido Rodrigues, Paulino~Ribeiro Villas~Boas, Lucas Antiqueira,
  Matheus~Palhares Viana, and Luis~Enrique Correa~Rocha.
\newblock Analyzing and modeling real-world phenomena with complex networks: a
  survey of applications.
\newblock {\em Advances in Physics}, 60(3):329--412, 2011.

\bibitem{da2006learning}
Luciano da~Fontoura~Costa.
\newblock Learning about knowledge: A complex network approach.
\newblock {\em Physical Review E}, 74(2):026103, 2006.

\bibitem{dacosta2020oncomplexity}
Luciano da~Fontoura~Costa.
\newblock On complexity and the prospects for scientific advancement.
\newblock {\em Revista Brasileira do Ensino de Física}, 2021.

\bibitem{dall2002random}
Jesper Dall and Michael Christensen.
\newblock Random geometric graphs.
\newblock {\em Physical review E}, 66(1):016121, 2002.

\bibitem{de2019connecting}
Henrique~F de~Arruda, Filipi~N Silva, Cesar~H Comin, Diego~R Amancio, and
  Luciano da~F Costa.
\newblock Connecting network science and information theory.
\newblock {\em Physica A: Statistical Mechanics and its Applications},
  515:641--648, 2019.

\bibitem{de2017knowledge}
Henrique~F de~Arruda, Filipi~N Silva, Luciano da~F Costa, and Diego~R Amancio.
\newblock Knowledge acquisition: A complex networks approach.
\newblock {\em Information Sciences}, 421:154--166, 2017.

\bibitem{foster2015tradition}
Jacob~G Foster, Andrey Rzhetsky, and James~A Evans.
\newblock Tradition and innovation in scientists’ research strategies.
\newblock {\em American Sociological Review}, 80(5):875--908, 2015.

\bibitem{lima2018dynamics}
Thales~S Lima, Henrique~F de~Arruda, Filipi~N Silva, Cesar~H Comin, Diego~R
  Amancio, and Luciano da~F Costa.
\newblock The dynamics of knowledge acquisition via self-learning in complex
  networks.
\newblock {\em Chaos: An Interdisciplinary Journal of Nonlinear Science},
  28(8):083106, 2018.

\bibitem{newman2003structure}
Mark~EJ Newman.
\newblock The structure and function of complex networks.
\newblock {\em SIAM review}, 45(2):167--256, 2003.

\bibitem{rzhetsky2015choosing}
Andrey Rzhetsky, Jacob~G Foster, Ian~T Foster, and James~A Evans.
\newblock Choosing experiments to accelerate collective discovery.
\newblock {\em Proceedings of the National Academy of Sciences},
  112(47):14569--14574, 2015.

\bibitem{shi2015weaving}
Feng Shi, Jacob~G Foster, and James~A Evans.
\newblock Weaving the fabric of science: Dynamic network models of science's
  unfolding structure.
\newblock {\em Social Networks}, 43:73--85, 2015.

\bibitem{silva2010identifying}
Filipi~Nascimento Silva, Bruno~AN Travencolo, Matheus~P Viana, and Luciano
  da~Fontoura~Costa.
\newblock Identifying the borders of mathematical knowledge.
\newblock {\em Journal of Physics A: Mathematical and Theoretical},
  43(32):325202, 2010.

\end{thebibliography}
\bibliographystyle{plain}

\end{document}